\documentclass[sigconf]{acmart}
\renewcommand\footnotetextcopyrightpermission[1]{} 
\usepackage{booktabs} 

\usepackage{multirow}
\usepackage{amsmath}
\usepackage{mathrsfs}

\usepackage{amsfonts,amssymb}
\usepackage{booktabs} 
\usepackage{multicol,color}
\usepackage{bm}
\usepackage{graphicx}
\usepackage{subfigure}
\usepackage{array}
\newcommand{\hide}[1]{} 
\newcommand{\vpara}[1]{\vspace{0.1in}\noindent\textbf{#1 }}

\newcommand{\smodel}{Progle\space}
\newcommand{\model}{Progle}

\usepackage[linesnumbered,ruled,vlined]{algorithm2e} 

\usepackage{float}
\begin{document}
\title{Spectral Network Embedding: A Fast and Scalable Method via Sparsity}

 \author{Jie Zhang, Yan Wang, Jie Tang, Ming Ding}
 \affiliation{
   \institution{Department of Computer Science and Technology, Tsinghua University}
 }
 \email{{j-z16,wang-y17, dm14}@mails.tsinghua.edu.cn,  jietang@tsinghua.edu.cn}

\begin{abstract}
Network embedding
aims to learn low-dimensional representations of nodes in a network, while the network structure and inherent properties are preserved. It has attracted tremendous attention recently due to significant progress in downstream network learning tasks, such as node classification, link prediction, and visualization. 
However, most existing network embedding methods suffer from the expensive computations due to the large volume of networks.
In this paper, we propose a $10\times \sim 100\times$ faster network embedding method, called \model, 
by elegantly utilizing the sparsity property of online networks and spectral analysis.
In \model, we first construct a \textit{sparse} proximity matrix and train the network embedding efficiently via sparse matrix decomposition. Then we introduce a network propagation pattern via spectral analysis to incorporate local and global structure information into the embedding.
Besides, this model can be generalized to integrate network information into other insufficiently trained embeddings at speed.
Benefiting from sparse spectral network embedding, our experiment on four different datasets shows that \smodel outperforms or is comparable to state-of-the-art unsupervised comparison approaches---DeepWalk, LINE, node2vec, GraRep, and HOPE, regarding accuracy, while is $10\times$ faster than the fastest word2vec-based method. Finally, we validate the scalability of \smodel both in real large-scale networks and multiple scales of synthetic networks.

\end{abstract}

%
%

\ccsdesc[500]{Information systems~Data mining}
\ccsdesc[300]{Information systems~Social networks}
\ccsdesc[500]{Computing methodologies~Dimensionality reduction and manifold learning}

 \keywords{network embedding, unsupervised learning, network spectral analysis, scalability}

\settopmatter{printacmref=false}
\setcopyright{none}
\maketitle

\section{Introduction}

Networks exist in many applications, e.g., social networks, gene-protein networks, road networks, and the World Wide Web.
One critical issue in network analysis is how to represent each node in the network. It is particularly challenging, as the network topology is often sophisticated and the scale of the network is extensive in many cases.

\hide{
-like structures occur in data in many guises, e.g., kinds of 
social networks, protein structure, gene regulatory networks and 
world-wide web. It is well recognized that network data is often sophisticated. To process these data effectively, the first critical challenge is to represent network concisely.
}

Network embedding, also called network representation learning, is a promising method to project nodes in a network to a low-dimensional continuous space while preserving some network properties. 
It could benefit a wide range of real applications such as link prediction, community detection, and node classification. Take the node classification as an example, and we can utilize network embedding to predict network node tags, such as interests of users in a social network, or functional labels of proteins in a protein-protein interaction network.

Roughly speaking, methodologies for network embedding fall into three categories: spectral embeddings, word2vec-based embeddings, and convolution-based embeddings. Spectral embedding approaches typically exploit the spectral properties of various matrix representations of graphs, especially the Laplacian and the adjacency matrices. Classical spectral methods can be viewed as dimensionality reduction techniques, and related works include  Isomap~\cite{tenenbaum2000global}, Laplacian Eigenmaps~\cite{belkin2001laplacian}, and spectral clustering~\cite{yan2009fast}. Involving matrix eigendecomposition, the expensive computational cost and poor scalability limit their application in real networks. 
 
\begin{figure*}[htbp]
\centering

\includegraphics[width = 1 \linewidth]{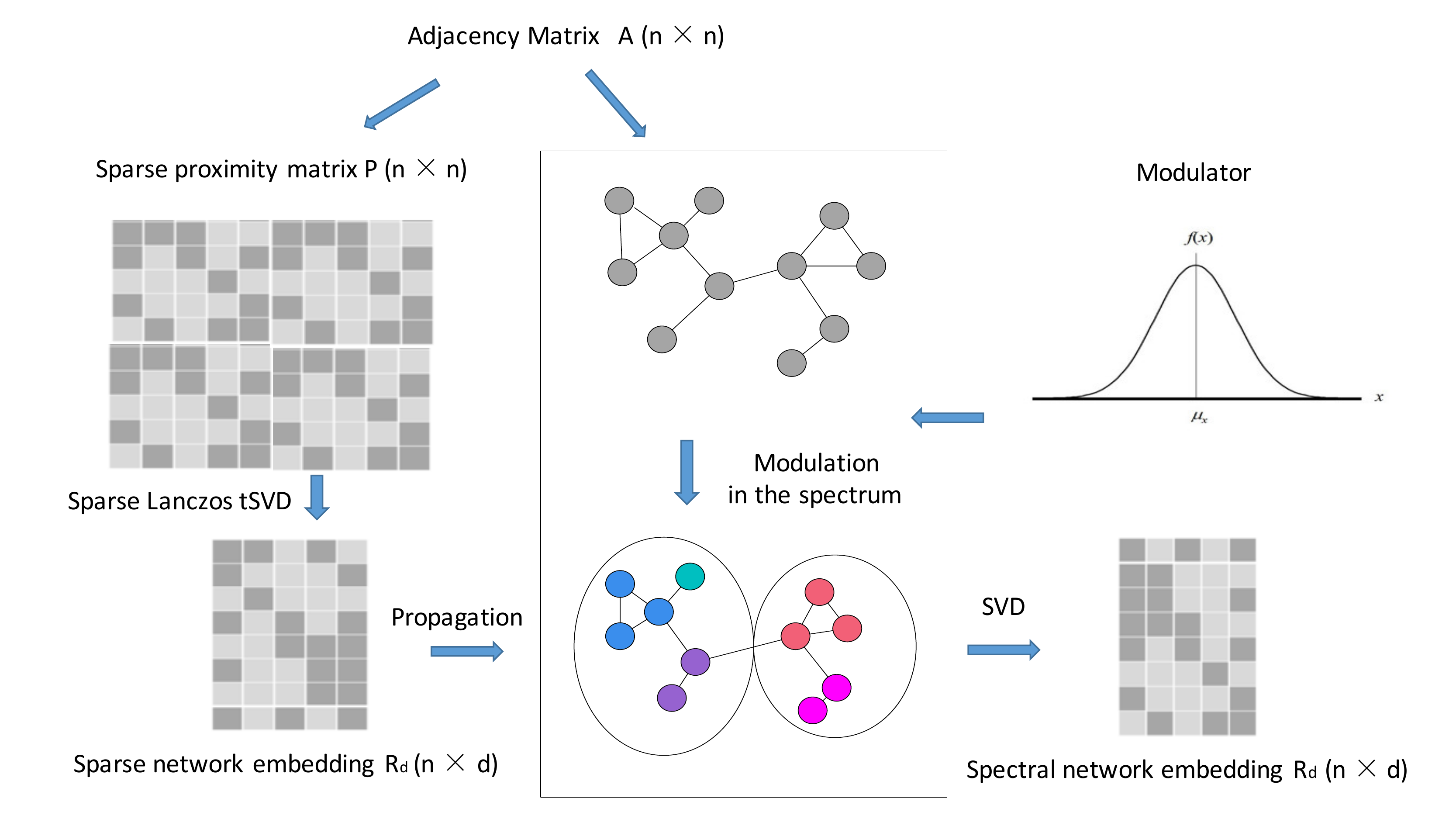}

\caption{\smodel Model: Given an adjacency matrix $A$, we first construct the sparse network proximity matrix $P$ via sparsity property; Secondly, get the raw network embedding via sparse Lanczos truncated SVD on the proximity matrix; Next, we modulate the network spectrum so that the network is locally smooth (spatially neighboring nodes in the network are dyed in the same color) and can be clustered well globally (the graph is partitioned into several subgraphs). Finally, we treat the raw embedding as a signal and propagate it in the modulated network to incorporate local and global network structure information.}
\label{fig:Pro}
\end{figure*}

\begin{figure}[htbp]
\hspace*{-0.8 \linewidth}
\centering

\includegraphics[width = 1.1\linewidth]{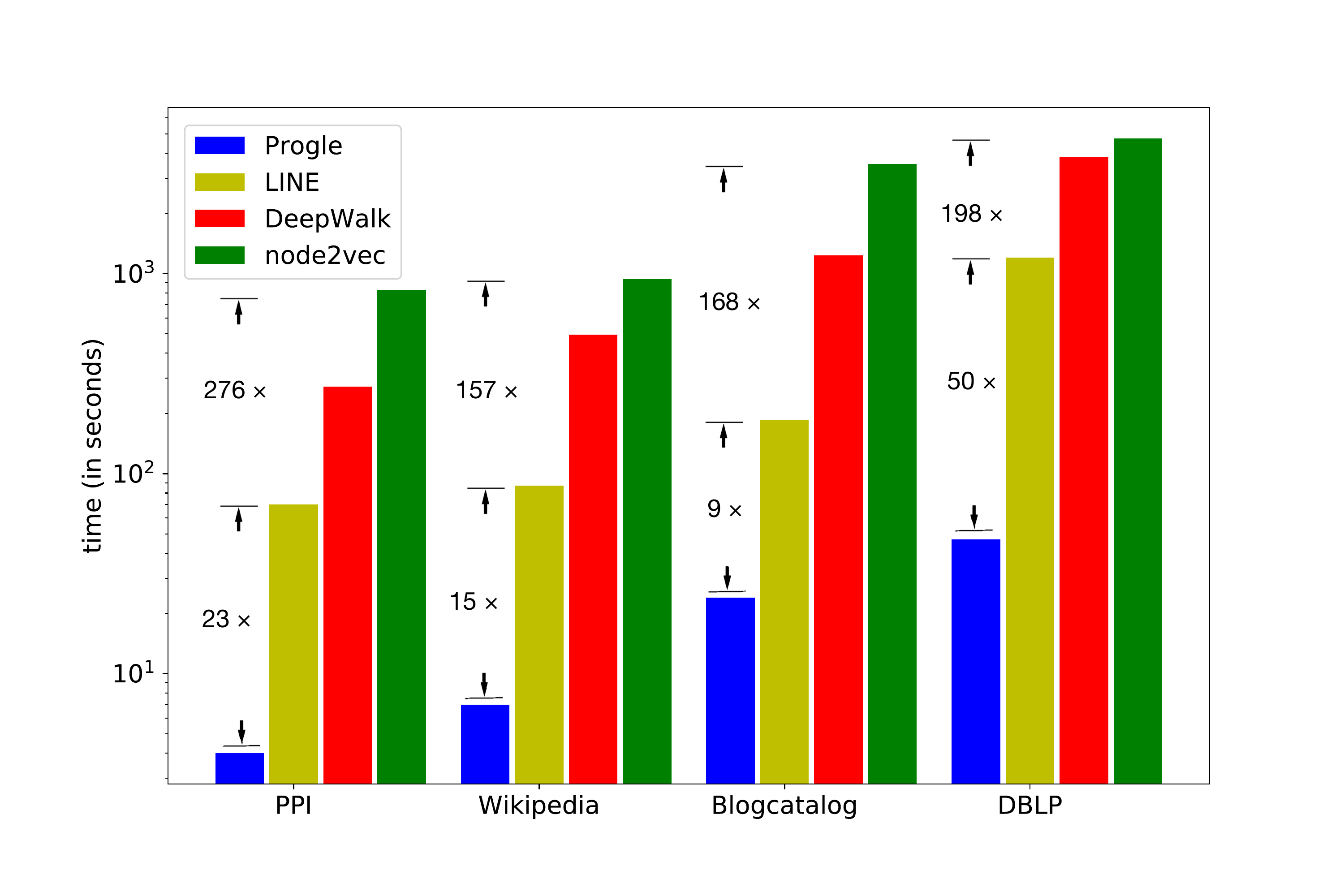}

\caption{Efficiency Comparison: Progle is $10 \times$ $\sim$ $100 \times$ faster than word2vec-based methods.}
\label{fig:Protime}
\end{figure}

Recently there has been a dramatic increase in the scalability of network embedding methods due to the introduction of the neural language model, word2vec~\cite{mikolov2013distributed}. 
The word2vec-based embedding methods, such as Deepwalk~\cite{perozzi2014deepwalk}, 
and node2vec~\cite{grover2016node2vec}, analogize nodes into words and capture network structure via random walks, which results in a large ``corpus" to train the node representations. They utilize SGD to optimize a neighborhood preserving likelihood objective that is usually non-convex. Though these methods are scalable and have achieved significant performance improvements in node classification and link prediction, they still suffer from high sampling time cost and optimization computation cost.

More recently, a few works have attempted to connect spectral embedding with  word2vec-based embedding.
Levy and Goldberg~\shortcite{levy2014neural} have proven that 
the neural word embedding is equal to implicit matrix factorization.
Following this thread, several matrix factorization embedding approaches using spectral dimensionality reduction techniques (e.g., SVD)
have been proposed, including
GraRep~\cite{cao2015grarep} and NetMF~\cite{qiu2017network}. 
These spectral methods demonstrated better statistical performances than original spectral approaches and corresponding word2vec-based counterparts.
However, they also inherit the defeats of the spectral methods, that is, the expensive computation and low scalability.


The third category of embedding method is convolution-based embedding method, for example, Graph Convolution Network (GCN)~\cite{kipf2016semi}, Graph Attention Network (GAT)~\cite{Velickovic:17GAT}, Message Passing Neural Networks (MPNNs)~\cite{gilmer2017neural} and GraphSage~\cite{hamilton2017inductive}.
The convolution-based embedding methods, as the supervised learning methods, further improve the embedding learning quality. Nevertheless, they need labeled data and are limited to specific tasks, and most of them are also not easily scaled up to handle large networks. More importantly, task-independent embedding learned in an unsupervised learning way often closely match task-specific supervised convolution-based approaches in predictive accuracy but benefit a broader range of other real applications. Therefore, we only discuss the unsupervised learning methods.

In summary, existing methods often suffer from the expensive computation or poor scalability. 
In this paper, inspired by the fact that most online networks follow power-laws and long-tailed distribution,  and the relationship between higher-order Cheeger's inequality and  spectral graph partitioning~\cite{lee2014multiway,bandeira2013cheeger},
we propose a scalable spectral embedding method, called \model,  with high efficiency and high accuracy.
In particular, we first construct the network proximity matrix via sparsity property. Secondly, it is deduced that the first phase network embedding (called sparse network embedding) in our model is decomposing a sparse matrix, and we can train network embeddings via sparse truncated Lanczos SVD algorithm. Finally, leveraging the relationship between higher-order Cheeger's inequality and multi-way graph partitioning, we modulate the network spectrum and get the final embedding (called spectral network embedding) mainly by propagating the sparse network embedding in the modulated network to incorporate local and global network information. 

The whole algorithm process, illustrated in Figure~\ref{fig:Pro}, only involves sparse matrix product and sparse spectral methods, and thus is scalable and \textbf{several orders of magnitude faster} than the common spectral embedding methods~(e.g., GraRep and NetMF). Compared with word2vec-based methods, 
Figure~\ref{fig:Protime} shows that \smodel also achieves significantly better efficiency performances on all the different datasets (\textbf{$\mathbf{10\times \sim 100\times}$ faster}). Moreover, regarding accuracy, \smodel outperforms significantly, or marginally in some case, 
all the comparison methods---DeepWalk, LINE, node2vec, GraRep, and HOPE.

\hide{
\vpara{Contribution.}Our main contributions are three-fold:
\begin{itemize}
    \item We leverage a spectral analysis method to propagate a raw proximity matrix in the network to form an informational proximity matrix. This model can be degenerated to interpret other embedding models and further be generalized to incorporate network information to other insufficiently learned embedding at speed.
    
    \item We extract from the dense node-context proximity matrix two sparse parts, the local part and the global part to avoid expensive computation and heavy memory burden, and deduce that network embedding in our model is to decompose a sparse matrix. 
        
    \item Via network sparsity, the proposed spectral method is light and fast without loss of performance, and can be scalable for large networks.

    \end{itemize}
}

\vpara{Organization.}The rest of the paper is organized as follows. Section 2 formulates the problem definition. Section 3 describes the model framework and algorithms. Section 4 presents the experimental results. Section 5 reviews related works and Section 6 concludes the paper.

\section{Problem Formulation}

Network embedding aims to project network nodes into a low-dimensional continuous vector space. The resultant embedding vectors
can be leveraged as features for various network learning tasks, such as node classification, link prediction, and community detection. 

For the convenience of narration, we introduce following notations.
%
Let $G$ be a connected network $G=(V, E)$, with $n$ nodes $V=\{v_1, v_2,...,v_n\}$, $|E|$ edges $E\subset V\times V$. We use $A$ to denote the adjacency matrix (binary or weighted), and $D$ to denote a degree matrix, with $D_{ii}= \sum_j A_{ij}$.

\begin{definition}
Node Proximity and Proximity Matrix: Node proximity defines a graph kernel $\mathbb{P}(u,v)$ mapping two nodes to a real value, which reflects the similarity between node $u$ and $v$ in geometry distance or network structure roles they play, e.g., the network hub and the structural hole~\cite{lou2013mining}. The corresponding kernel matrix is called the proximity matrix $\mathbb{P}$.  
\end{definition}

%
\begin{definition}
Network Embedding: Given a network $G=(V, E)$, network embedding is a mapping function $f: V \mapsto R^d $ from network space to $d$-dimension space, where $d \ll |V|$. In the $R^d$ space, the distance $d(u, v)$ characterizes some proximity between node $u$ and $v$ in the original network space so that embedding vector set can be utilized as features for machine learning in the network.
	
\end{definition}

By convention, the downstream learning tasks are reversely utilized to evaluate the quality of the learned embedding. Generally, there are two goals for network embedding, the original network reconstruction from the learned embedding and supporting of network inference~\cite{cui2017survey}. The limitation of former is stated in~\cite{cui2017survey}: the embedding may overfit the adjacency matrix. We use the standard method used in Deepwalk to validate the embedding's inference ability, multi-label classification task, where each node may have multiple labels to infer~\cite{tang2009large, perozzi2014deepwalk}. All comparisons involved are unsupervised learning algorithms.


\section{Model Framework}

In this section, we first leverage the sparse property of networks to learn sparse network embedding, then incorporate local and global network information into the embedding via spectral methods. In our model, it' s obvious to conclude that both time complexity and space complexity are linear to the volume of network and the proposed approach is efficient and scalable for large networks.

\subsection{Sparse Network Embedding}
Since our method is based on the spectral method, we first analyze the cause of expensive computation cost and poor scalability of state-of-the-art spectral matrix factorization embedding methods. 
\subsubsection{Cause of Expensive Computation Cost and Poor Scalability}
The notable word embedding method, word2vec, learning distributional representations for words, is based on the distribution hypothesis\cite{harris1954distributional}, which states words in similar contexts have similar meanings. Similarly, the word2vec-based embedding methods~(Deepwalk, node2vec, LINE, etc.) and some spectral methods (GraRep, NetMF, etc.) transfer this hypothesis to networks and assume that nodes in similar network contexts are similar. In general, the contexts of a node are defined as the node set it can arrive within $m$ steps. These methods define the following node-context proximity matrix implicitly or explicitly~\cite{levy2014neural,qiu2017network}:
 \begin{align}
\label{equ:p1}
\mathbb{P} = \frac{P^1 + P^2 +...+P^m}{m}
\end{align}
where $P=D^{-1}A$ is called transmission matrix and the default value of $m$ is $10$. The entity $\mathbb{P}_{ij}$ reflects the probability proximity value between node $i$ and it's context node $j$. The objective of state-of-the-art spectral embedding methods is to use a function of dot product of node embedding $r_i$ and context node embedding $c_j$ to approximate $\mathbb{P}_{i,j}$. 

The equation~(\ref{equ:p1}) is also inherited from the linear bag-of-word context assumption in skip-gram model\cite{mikolov2013distributed} that context words in a size $m$ window of the target word share the same weight. Although equation~(\ref{equ:p1}) is expressive to learn word embedding for the sequential property of the sentences, it still ignores spatial locality and sparsity of the network, resulting in the dense matrix $\mathbb{P}$. Since these matrix factorization methods are to factorize a matrix derived from $\mathbb{P}$, the dense matrix $\mathbb{P}$ is a cause of expensive computation cost and poor scalability, with $O(|V|^3)$ time complexity and $O(|V|^2)$ storage space complexity.

Even if we make a trade-off between efficiency and effectiveness and set $m=1$, the objective matrix is still dense. The denseness of the objective factorized matrix also comes from negative sampling tricks\cite{mikolov2013distributed}, resulting the factorized matrix is in the form of a shifted PMI matrix~\cite{church1990word}, with a global shifted bias $-\log k$ ($k$ is the negative sampling number). Even all negative values in the objective matrix are replaced by $0$ ~\cite{levy2014neural, cao2015grarep}, but it cannot reduce the magnitude order of non-zero entities.

To deal with the causes of expensive computation cost and poor scalability, we first give a reasonable sparse proximity matrix and restrict our mathematical derivation in the sparse node-context edge set and conclude our embedding can be achieved by sparse spectral matrix decomposition. In the Spectral Propagation section, we further eliminate the possible loss of accuracy and the model's expression capacity resulting from sparse processing. The final experiments will support our model.

\subsubsection{Sparse Proximity Matrix}
 The transmission matrix $P$ to the power of different orders in the equation~(\ref{equ:p1}) reflects different order proximities between nodes and context nodes. Since the network is nonlinear, the weights of different order proximities are inconsistent with the linear bag-of-word model. Intuitively, lower order proximities reveal the basic connectivity of networks hence play the main part of proximity matrix. In GraRep~\cite{cao2015grarep}, different order proximities are considered separately. In HOPE~\cite{ou2016asymmetric}, Katz matrix is used to replace the equation~(\ref{equ:p1}),
 \begin{align}
\label{equ:karz}
\mathbb{S}^{Karz} = \sum_{i=1}^{\infty}\beta A^i
\end{align}
where $\beta$ is a decay parameter. 

Here we use an efficient edge dropout method to give an attenuation sum of different order proximities while remaining the resultant sum sparse. We draw the edges of adjacency matrix with a dropout ratio $\eta$ to achieve a series of $\{\hat{A}_i\}_{i=1}^{m-1}$. We replace the non-zero elements of a matrix with $1$ and denote the operation as $\langle{.}\rangle$. Our sparse proximity matrix is
 \begin{align}
\label{equ:ourp}
\mathbb{P} \sim \sum_{i=1}^{m} (D^{-1}A)^i \circ \langle{A\prod_{j=1}^{i-1} \hat{A}_j}\rangle
\end{align}
where $\circ$ is Hadamard product, also named element-wise product. $\langle{A\prod_{j=1}^{i-1} \hat{A}_j}\rangle
$ controls the sparsity and plays the role of a decay factor. $m$ only needs to be set to $1, 2, 3$, as the possible loss can be compensated later.

After normalizing the right hand of the equation~(\ref{equ:ourp}), we get the sparse proximity matrix, still with high order proximities. The whole process only involves sparse matrix product and is efficient.


\subsubsection{Network Embedding as Sparse Matrix Decomposition} The edges of the sparse proximity matrix form a node-context pair set $\mathcal{D}$. We define the occurrence probability of context $j$ given node $i$ as
\begin{align}
\label{equ:pro}
 \hat{p}_{i,j} = \sigma(c_j^Tr_i)
\end{align}
where $\sigma(x) = 1/(1+e^{-x})$ is the sigmoid function, $c_j$ is the the context vector of context node $j$ and $r_i$ is the representation embedding vector of node $i$. The objective function can be expressed as the weighted sum of log loss:
\begin{align}
\label{equ:loss0}
l = -\sum_{(i,j)\in \mathcal{D}} p_{i,j}\ln \hat{p}_{i,j} 
\end{align}
where $p_{ij}$ is the entity in proximity matrix $\mathbb{P}$ and indicates the weight of $(i,j)$ in the node-context pair set $\mathcal{D}$. 

However, this objective admits a trivial solution in which $r_i=c_j$, $c_j^Tr_i = K$($K$ is very large) then $\hat{p}_{i,j}=1$ for every pair $(i,j)$, resulting in the consequence that all the node embeddings are the same and cannot represent any network information about the nodes.

To prevent this trivial solution, we suppose that given an observed node-context pair $(i,j)$, the appearance of context $j$ is caused by $i$'s stimulation,  or by the background 
noise $P_{\mathcal{D},j}$ of the node-context set $\mathcal{D}$. The binary classifier $\hat{p}_{i,j}$ should distinguish the noise $j$ as a negative sample. So the modified loss is:
\begin{align}
\label{equ:loss}
l = -\sum_{(i,j) \in \mathcal{D}} [p_{i,j}\ln \sigma (c_j^T r_i) 
+  \lambda P_{\mathcal{D},j}\ln \sigma (-c_j^T r_i)]
\end{align}
where $\lambda$ is the negative noise sample ratio. $P_{\mathcal{D},j}$, the background noise of the node-context set $\mathcal{D}$ associated with context node $j$, can be defined below:
\begin{align}
\label{equ:pd}
P_{\mathcal{D},j} = \frac{\sum_{i:(i,j)\in \mathcal{D}} p_{i,j}}{\sum_{(i,j)\in \mathcal{D}} p_{i,j}} 
\end{align}

A sufficient condition for minimizing the objective (\ref{equ:loss}) is to let its partial derivative with respect to $c_j^T r_i$ be zero. We can get:
\begin{align}
\label{equ:rc}
r_i^T c_j = \ln p_{i,j} - \ln (\lambda P_{D,j}), \quad (i,j)\in \mathcal{D}
\end{align}

We define a matrix $M$ whose entities are given below:
\begin{align}
\label{equ:mm}
M_{i,j} = \left\{
\begin{aligned}
	&\ln p^l_{i,j} - \ln (\lambda P_{D,j}) &, (i,j)\in \mathcal{D}\\
&0 &, (i,j)\notin \mathcal{D}
\end{aligned}
\right.
\end{align}

From equation (\ref{equ:mm}), the objective is transformed to approximate the matrix $M$ with the row-rank product of representation embedding matrix $R$ and context embedding matrix $C$. An alternative optimization method is truncated Singular Value Decomposition (tSVD), which achieves the optimal rank $d$ factorization regarding $L_2$ loss. The objective is shown below:
\begin{align}
\label{equ:Pro}
\min_{R_d,C_d} ||M - RC^T||_F
\end{align}
where $R_d$ and $C_d$ are $n \times d$ matrices whose row stands for a node's embedding and context embedding respectively. The normalized $R_d$ is:
 \begin{align}
\label{equ:sy}
R_d \leftarrow R_d \Sigma_d^{1/2}
\end{align}

Benefiting from the sparse proximity matrix construction, $|\mathcal{D}| \ll |V \times V|$, sparse truncated Lanczos SVD can be utilized on matrix $M$ with time complexity $O(|E|d^2)$.

%

\subsection{Spectral Propagation} 
In this section, we introduce a propagation way via spectral analysis and interpret the intuition why spectral propagation can incorporate the global clustering information and spatial locality smoothing into the sparse embedding. Hence it further eliminates the possible loss of accuracy or expression capacity resulting from sparse processing. Moreover, spectral propagation is also a general embedding enhancement method.

\subsubsection{Network Laplacian Transform}


Unnormalized network Laplacian is an essential operator in spectral network analysis and defined as $L=D-A$. The random walk normalized Laplacian definition is $L = E_n-D^{-1}A$, where $E_n$ is the  identity matrix. The random walk normalized Laplacian can be decomposed as $L = U\Lambda U^T$, where $\Lambda = diag([\lambda_1,...,\lambda_n])$ with $0=\lambda_1 \leq ... \leq\lambda_n$ and $U$ is the square($n \times n$) matrix whose $i^{th}$ column is the eigenvector $u_i$.

The complete set of orthonormal eigenvectors $\{u_i\}_{i=1}^n$ are also known as the graph Fourier modes, and the associated eigenvalues $\{u_i\}_{i=1}^n$ are identified as the frequencies of the graph. The graph Fourier transform of a signal $x$ is defined as $\hat{x} = U^T x$ while the inverse transform as $x= U\hat{x}$. They are the transforms between original ``temporal'' space and spectral (frequency) space. A simple network propagation $D^{-1}Ax$ can be interpreted that the signal $x$ is first transformed into the spectral space and scaled by the eigenvalues, then transformed back.

We will demonstrate next that the eigenvalues in the spectral space are closely associate with networks' spatial locality smoothing and global clustering, according to higher-order Cheeger' inequality~\cite{lee2014multiway,bandeira2013cheeger}. 

\subsubsection{Multi-way Graph Partitioning and Higher-order Cheeger's Inequality}

\begin{figure}{
\centering
\includegraphics[width = 0.7\linewidth]{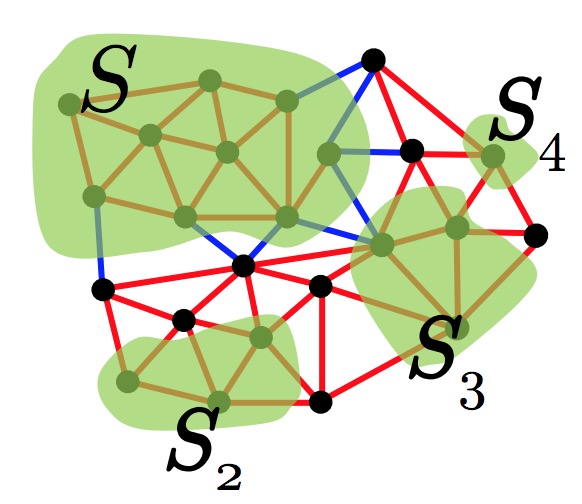}
\caption{Expansion and $k$-way Cheeger constant. The expansion $\phi (S)$ indicates the effect of the graph partitioned by $S$. Its value is related to the blue edges shown in the figure. $k$-way Cheeger constant reflects the effect of the graph partitioned into $k$ parts and $k=4$ here.}
\label{fig:cheeger}
}
\end{figure}

We first introduce the definitions of expansion and $k$-way Cheeger constant.
\begin{definition}
Expansion: For a node subset $S \subseteq V$
$$\phi(S) = \frac{|E(S)|}{min\{vol(S), vol(V-S)\}}$$
where E(S) is the set of edges with one point in $S$ and vol($S$) is the sum of nodes' degree in set $S$.
\end{definition}

\begin{definition}
$k$-way Cheeger constant: $$\rho_G(k) = min \{max\{\phi(S_i): S_1, S_2,..., S_k \subseteq V ~disjoint\}\} $$
\end{definition}

According to the definitions, the expansion indicates the effect of the graph partitioned by a subset and
$k$-way Cheeger constant reflects the effect of the graph partitioned into $k$ parts. Smaller value means better partitioning effect. An example is illustrated in figure~\ref{fig:cheeger}.

Higher-order Cheeger's inequality~\cite{lee2014multiway,bandeira2013cheeger} bridges the gap between network spectral analysis and graph partitioning via controling the bounds of $k$-way Cheeger constant:
\begin{align}
\label{equ:cheeger}
\frac{\lambda_k}{2}\leq \rho_G(k) \leq O(k^{2})\sqrt{\lambda_k}
\end{align}

A basic fact in spectral graph theory is that the number of connected components in an undirected graph is equal to the multiplicity of the eigenvalue zero in the Laplacian matrix, which can be concluded from $\rho_G(k)=0$ when setting $\lambda_k =0$ in inequality~(\ref{equ:cheeger}).

From inequality~(\ref{equ:cheeger}), we can conclude that small (large) eigenvalues control global clustering (local smoothing) effect of the network partitioned into a few large parts (many small parts). This inspires us to incorporate global and local network information into network embedding when propagating the raw embedding in the modulated network. For the convenience of narration, we discuss the modulation of the Laplacian matrix and get the new embedding via 
\begin{align}
\label{equ:DAR}
R_d \gets \widetilde{D^{-1}A}R_d = (E_n - \widetilde{L})R_d
\end{align}

Let $\widetilde{L} = Ug(\Lambda)U^T $ where $g$ is the  modulator in the spectrum.

To take both local and global structure information into consideration, we design the spectral modulator as $g(\lambda) = e^{-\frac{1}{2}[(\lambda-\mu)^2-1] \theta}$, where $\mu \in [0,2]$. Therefore, 
\begin{align}
\label{equ:LUU}
\widetilde{L} &= U diag([e^{-\frac{1}{2}[(\lambda_1-\mu)^2-1] \theta},..., e^{-\frac{1}{2}[(\lambda_n-\mu)^2 -1]\theta}])U^T 
\end{align}

 $g(\lambda)$ is a band-pass filter kernel~\cite{shuman2016vertex, hammond2011wavelets} that passes eigenvalues within a certain range and attenuates eigenvalues outside that range, hence $\rho_G(k)$ is attenuated for corresponding top largest eigenvalues and top smallest eigenvalues, resulting in the amplified global and local network information. A band-pass graph modulation example is illustrated in Figure~\ref{fig:Pro}. As for the low-pass or high-pass filter, which only amplifies the local or global information respectively, can be achieved by tuning the position parameter $\mu$ of the band-pass filter. Therefore, the band-pass filter is an general spectral network modulator.

\subsubsection{Chebyshev Expansion}

Note that the Fourier transform and inverse Fourier transform in equation~(\ref{equ:LUU}) involve eigendecomposition, we utilize truncated Chebyshev expansion to avoid explicitly transforming. The Chebyshev polynomials of the first kind are defined by the recurrence relation $T_{i+1}(x)=2xT_i(x)-T_{i-1}(x)$ with $T_0(x)=1, T_1(x)=x$.
 Then
 \begin{align}
\label{equ:chebyshev1}
\widetilde{L} \approx U\sum_{i=0}^{k-1}c_i(\theta)T_i(\bar{\Lambda})U^T = \sum_{i=0}^{k-1}c_i(\theta)T_i(\bar{L})
\end{align}
where $\bar{\Lambda} = -\frac{1}{2}[(\Lambda-\mu E_n)^2-E_n]$, $\bar{L} = -\frac{1}{2}[(L-\mu E_n)^2-E_n]$ and scaled eigenvalues lie in $[-1,1]$. The new kernel is $f(\bar{\lambda}) = g(\lambda) = e^{-\bar{\lambda} \theta}$. 

As $T_i$ are orthogonal with the weight $1/\sqrt{1-x^2}$ on the interval $[-1,1]$, that is,
 \begin{align}
\label{equ:orthogonal}
\int_{-1}^{1}\frac{T_i(x)T_j(x)}{\sqrt{1-x^2}}dx = 
\left\{
\begin{aligned}
	&0 &, i \ne j\\
	&\pi &, i = j = 0\\
    &\frac{\pi}{2} &, i = j \ne 0
\end{aligned}
\right.
\end{align}
 the coefficients of Chebyshev expansion for $e^{-x \theta}$ can be get by:
\begin{align}
\label{equ:chebyshev2}
c_i(\theta) = 
\left\{
\begin{aligned}
	& \frac{1}{\pi}\int_{-1}^{1}\frac{T_i(x)e^{-x \theta}}{\sqrt{1-x^2}}dx = (-)^iI_i(\theta) &, i = 0\\
	& \frac{2}{\pi}\int_{-1}^{1}\frac{T_i(x)e^{-x \theta}}{\sqrt{1-x^2}}dx = 2(-)^iI_i(\theta) &, i \ne 0
\end{aligned}
\right.
\end{align}
where $I_i(\theta)$ is the notable special function, modified Bessel function of the first kind~\cite{andrews1992special}.

Then, we get the series expansion of the modulated Laplacian:
 \begin{align}
\label{equ:chebyshev1}
\widetilde{L} \approx I_0(\theta)T_0(\bar{L}) + 2\sum_{i=1}^{k-1}(-)^iI_i(\theta)T_i(\bar{L})
\end{align}

The embedding matrix propagated by the spectral modulated network is:

  \begin{align}
  \label{equ:rd1}
R_d &\gets \widetilde{D^{-1}A}R_d = (E_n - \widetilde{L})R_d \notag \\
&= \{E_n - [I_0(\theta)T_0(\bar{L}) + 2\sum_{i=1}^{k-1}(-)^iI_i(\theta)T_i(\bar{L})]\} R_d
\end{align}

To alleviate the perturbation of the spectrum of $D^{-1}A$, we further modify the expression~(\ref{equ:rd1}):

  \begin{align}
  \label{equ:rd2}
R_d &\gets D^{-1}A \widetilde{D^{-1}A}R_d = D^{-1}A(E_n - \bar{L})R_d \notag \\
&= D^{-1}A \{E_n - [I_0(\theta)T_0(\bar{L}) + 2\sum_{i=1}^{k-1}(-)^iI_i(\theta)T_i(\bar{L})]\} R_d
\end{align}

The computation of expression~(\ref{equ:rd2}) can be efficiently executed in a recurrence way.
Denote $\bar{R}^{(i)}_d=T_i(\bar{L}) R_d$, then $\bar{R}^{(i)}_d=2\bar{L}\bar{R}^{(i-1)}_d-\bar{R}^{(i-2)}_d$ with $\bar{R}^{(0)}_d=R_d$ and $\bar{R}^{(1)}_d=\bar{L}R_d$. Note that $\bar{L} = -\frac{1}{2}[(L-\mu E_n)^2-E_n]$ and $L$ is sparse, the overall complexity of equation~(\ref{equ:rd2}) is $O(k|E|)$.

In addition, while the spectral propagation incorporates more structure information into embedding, it also ruin the orthogonality of original embedding space achieved by sparse SVD. Hence, we mend it via common SVD on $R_d$. As $R_d$ is $n \times d$, the time complexity is $O(|V|d^2)$.

\subsubsection{Spectral Propagation as a General Enhancement Method}
Though we propose spectral propagation to eliminate the possible loss of accuracy due to sparse processing, the spectral propagation is somewhat independent of sparse network embedding model. In other word, it is a general method of further incorporating network information efficiently and thus enhance the quality of embedding trained by other models. We will validate the effectiveness of enhancement in the experiment part.

\subsection{Algorithm summary}
In this part, we summarize the proposed algorithm, \model. We firstly construct the network proximity matrix via sparsity property; Secondly we deduce that the raw network embedding in our model is decomposing a sparse matrix; Next, we can train network \textbf{E}mbeddings via sparse truncated Lanczos SVD algorithm; Finally, according to higher-order Cheeger's inequality, we modulate the network's spectrum and and \textbf{Pro}pagate the raw embedding in the whole modulated network to incorporate \textbf{G}lobal and \textbf{L}ocal network structure information. 

Without the product and decomposition operations on the dense matrix,
the proposed method can be scalable for large networks efficiently with competitive performance and takes up fewer computation resources due to sparsity~\cite{berry1991multiprocessor,berry1992large}.

\subsubsection{Complexity analysis}
In the sparse network embedding part, the sparse proximity matrix generation costs $O(|E|)$ time and the sparse network embedding training (sparsity truncated SVD) costs $O(|E|d^2)$. In the spectral propagation part, the propagtion complexity is $O(k|E|)$ and the final SVD costs $O(|V|d^2)$. In all, the total time complexity is $O((|V|+|E|)d^2)$.

The space complexity to restore the sparse proximity matrix and embedding is $O(|V|d+|E|)$. 

Both of time complexity and space complexity are linear to the volume of network, which accounts for the efficiency and scalability.

\section{Experiments}
In this section, we conduct the multi-label node classification experiments to evaluate the proposed model and the improvement for other embedding methods. Then we compare our time cost with the baselines, and further validate \model 's scalability and efficiency in large-scale real networks and multiple scales of synthetic networks.

\subsection{Experimental Setup}
\subsubsection{Datasets}
We employ four widely-used datasets for the node classification experiments. The statistics are shown in Table~\ref{tab:data}.
\begin{table}[htbp]
\centering
	\caption{Labeled data statistics.}
	\label{tab:data}
	\begin{tabular}{c c c c c}
\toprule[1.1pt]
		\textit{Dataset} & \textit{Blogcatalog} & \textit{Wikipidia} & \textit{PPI} & \textit{DBLP} \\
		\hline
		\#\textit{Nodes} & 10,312 & 4,777 & 3,890 & 51,264 \\
		\#\textit{Edges} & 333,983 & 184,812 & 76,584 & 127,968 \\
		\#\textit{Labels} & 39 & 40 & 50 & 60 \\
		\bottomrule[1.1pt]
	\end{tabular}
\end{table}

\begin{itemize}
	\item BlogCatalog~\cite{zafarani2009social}: BlogCatalog is a social blogger network, where nodes and edges stand for bloggers and their pairwise social relationships, respectively. Bloggers submit their blogs with interest tags, which are considered as ground-truth labels.
	\item Wikipedia~\cite{mahoney2009large}: This is a co-occurrence network of words in the first million bytes of the Wikipedia dump, and the labels represent the Part-of-Speech(POS) tags. 
	\item Protein-Protein Interactions (PPI)~\cite{breitkreutz2008biogrid}: This is a subgraph of PPI network for Homo Sapiens. The subgraph corresponds to the graph induced by nodes which have labels from hallmark gene sets and represent biological states.
	\item DBLP\cite{tang2008arnetminer}: DBLP is an academic network dataset where authors are treated as nodes, citation relationships as edges and academic conferences or activities as labels.

\end{itemize}

Another three real large-scale networks without labels, and multiple scales of synthetic networks are added to validate \model 's efficiency and scalability.
\begin{table}[htbp]
\centering
	\caption{Unlabeled data statistics.}
	\label{tab:data2}
	\begin{tabular}{c c c c c}
\toprule[1.1pt]
		\textit{Dataset} &\textit{Synthetic} &\textit{Flickr} & \textit{Youtube} & \textit{wiki-topcats}\\
		\hline
		\#\textit{Nodes} &$-$ &80,513 & 1,134,890 & 1,791,489 \\
		\#\textit{Edges} &$-$ &5,899,882 & 2,987,624 & 25,447,873 \\
		\bottomrule[1.1pt]
	\end{tabular}
\end{table}

\begin{table*}[htbp]
\centering
     \caption{Micro-F1 of multi-label classification on different datasets}
	\label{tab:Random Walk}
     \begin{tabular}{c|c|ccccccccc}
		\toprule[1.1pt]
		Dataset &training ratio  & 0.1 & 0.2 & 0.3 & 0.4 & 0.5 & 0.6 & 0.7 & 0.8 & 0.9 \\
		\hline
		\multirow{5}{*}{PPI} 
	    &DeepWalk  & 0.164 & 0.185 & 0.194 & 0.203 & 0.211 & 0.218 & 0.223 & 0.226 & 0.227 \\
		&LINE  & 0.163 & 0.189 & 0.201 & 0.210 & 0.215 & 0.221 & 0.227 & 0.229 & 0.231 \\ 
	    &Node2vec  & 0.162 & 0.184 & 0.197 & 0.204 & 0.216 & 0.222 & 0.231 & 0.231 & 0.241 \\
  		 &GraRep & 0.154 & 0.176 & 0.189 & 0.199 & 0.202& 0.205 & 0.204 & 0.207 & 0.209  \\
	     &HOPE & 0.164 & 0.186 & 0.198 & 0.206 & 0.210& 0.215 & 0.217 & 0.222 & 0.225  \\
	    &Progle & \textbf{0.182} & \textbf{0.212} & \textbf{0.227} & \textbf{0.237} & \textbf{0.246} & \textbf{0.249} & \textbf{0.254} & \textbf{0.258} & \textbf{0.259}  \\
	    & ($\pm\sigma$) &($\pm$0.005) & ($\pm$0.004) & ($\pm$0.003) & ($\pm$0.006) & ($\pm$0.007) & ($\pm$0.009) & ($\pm$0.010) & ($\pm$0.010)  & ($\pm$0.011)\\
		\hline
	\multirow{5}{*}{Wikipedia} 
&DeepWalk  & 0.404 & 0.431 & 0.459 & 0.477 & 0.485 & 0.487 & 0.491 & 0.492 & 0.494 \\
		&LINE  & \textbf{0.478} & 0.496 & 0.504 & 0.510 & 0.512 & 0.516 & 0.516 & 0.517 & 0.524 \\ 
		&Node2vec  & 0.456 & 0.463 & 0.470 & 0.472 & 0.482 & 0.487 & 0.496 & 0.498 & 0.500 \\
	    &GraRep & 0.472 & 0.488 & 0.497 & 0.504 & 0.506& 0.507 & 0.509 & 0.510 & 0.518  \\
	     &HOPE & 0.385 & 0.395 & 0.398 & 0.402 & 0.401& 0.402 & 0.401 & 0.398 & 0.401  \\
	    &Progle & 0.473 & \textbf{0.513} & \textbf{0.531} & \textbf{0.538} & \textbf{0.547} & \textbf{0.550} & \textbf{0.552} & \textbf{0.557} & \textbf{0.572}  \\
	        & ($\pm\sigma$) &($\pm$0.007) & ($\pm$0.005) & ($\pm$0.004) & ($\pm$0.004) & ($\pm$0.008) & ($\pm$0.004) & ($\pm$0.008) & ($\pm$0.012)  &($\pm$0.013)\\
		\hline
	\multirow{5}{*}{Blogcatalog} 
		&DeepWalk  & 0.362 & 0.384 & 0.396 & 0.405 & 0.409 & 0.411 & 0.414 & 0.417 & 0.422  \\
		&LINE  & 0.282 & 0.299 & 0.306 & 0.310 & 0.332 & 0.345 & 0.355 & 0.360 & 0.368 \\ 
	    &Node2vec  & \textbf{0.363} & 0.385 & 0.397 & \textbf{0.408} & 0.411 & 0.417 & 0.420 & 0.422 & 0.421 \\
	    &GraRep & 0.340 & 0.320 & 0.325 & 0.330 & 0.333& 0.336 & 0.337 & 0.338 & 0.341  \\
	     &HOPE & 0.307 & 0.326 & 0.334 & 0.339 & 0.343& 0.346 & 0.350 & 0.353 & 0.353  \\
	    &Progle & 0.362 & \textbf{0.388} & \textbf{0.400} & 0.407 & \textbf{0.412} & \textbf{0.418} & \textbf{0.421} & \textbf{0.426} & \textbf{0.427}  \\
	    & ($\pm\sigma)$ &($\pm$0.005) & ($\pm$0.004) & ($\pm$0.003) & ($\pm$0.004) & ($\pm$0.006) & ($\pm$0.007) & ($\pm$0.007) & ($\pm$0.009)  &($\pm$0.012)\\
     \bottomrule[1.1pt]
     		Dataset &training ratio  & 0.01 & 0.02 & 0.03 & 0.04 & 0.05 & 0.06 & 0.07 & 0.08 & 0.09 \\
		\hline
     	\multirow{5}{*}{DBLP} 
			&DeepWalk  & 0.493 & 0.532 & 0.550 & 0.562 & 0.571 & 0.575 & 0.579 & 0.582 & 0.584 \\
		&LINE  & 0.487 & 0.514 & 0.526 & 0.532 & 0.535 & 0.538 & 0.541 & 0.544 & 0.545 \\ 
	    &Node2vec  & 0.489 & 0.532 & 0.551 & 0.563 & 0.570 & 0.576 & 0.580 & 0.582 & 0.584 \\
	    &GraRep & 0.505 & 0.521 & 0.526 & 0.529 & 0.532 & 0.534 & 0.535 & 0.537 & 0.538  \\
	     &HOPE & \textbf{0.522} & \textbf{0.542} & 0.550 & 0.555 & 0.559 & 0.561& 0.563 & 0.565  & 0.566  \\
	    &Progle  & 0.488 & \textbf{0.542} & \textbf{0.562} & \textbf{0.573} & \textbf{0.580} & \textbf{0.584} & \textbf{0.588} & \textbf{0.590} & \textbf{0.592}  \\
    & ($\pm\sigma$) & ($\pm$0.010) & ($\pm$0.006) & ($\pm$0.005) & ($\pm$0.004) & ($\pm$0.002) & ($\pm$0.002) & ($\pm$0.002) & ($\pm$0.002)  & ($\pm$0.001) \\
     \bottomrule[1.1pt]
  	 \end{tabular} 
\end{table*}

\begin{itemize}
\item Synthetic networks: They are regular graphs. The network degree ranges from 10 to 1000, and the number of node ranges from 1000 to 10,000,000.
\item Flickr~\cite{zafarani2009social}: A user contact network dataset crawled from Flickr. 
\item Youtube~\cite{leskovec2016snap}: 
     A friendship network of Youtube users. 
\item wiki-topcats~\cite{leskovec2016snap}: 
A web graph of Wikipedia hyperlinks collected in September 2011.
\end{itemize}

\subsubsection{Comparison baselines}
The following methods are baselines in the experiments:
\begin{itemize}
 	\item DeepWalk\cite{perozzi2014deepwalk}. DeepWalk transforms a graph structure into linear sequences by random walks and processes the sequences using word2vec model (skip-gram).
 		
 	\item LINE\cite{tang2015line}. It defines loss functions to preserve first-order or second-order proximity in the embedding. 
    
 	\item node2vec~\cite{grover2016node2vec}. It can be treated as the biased random walk version of DeepWalk.
 	\item GraRep~\cite{cao2015grarep}. It decomposes $k$-step probability transition matrix to train the node embedding, then concatenate all $k$-step representations.
 	
 	\item HOPE~\cite{ou2016asymmetric}. It is a matrix factorization spectral method, approximating high-order proximity based on factorizing the Katz matrix.
\end{itemize}

\begin{figure*}[htbp]
\centering
\begin{subfigure}[ProDeepWalk]{
\centering
\includegraphics[width = 0.4 \linewidth]{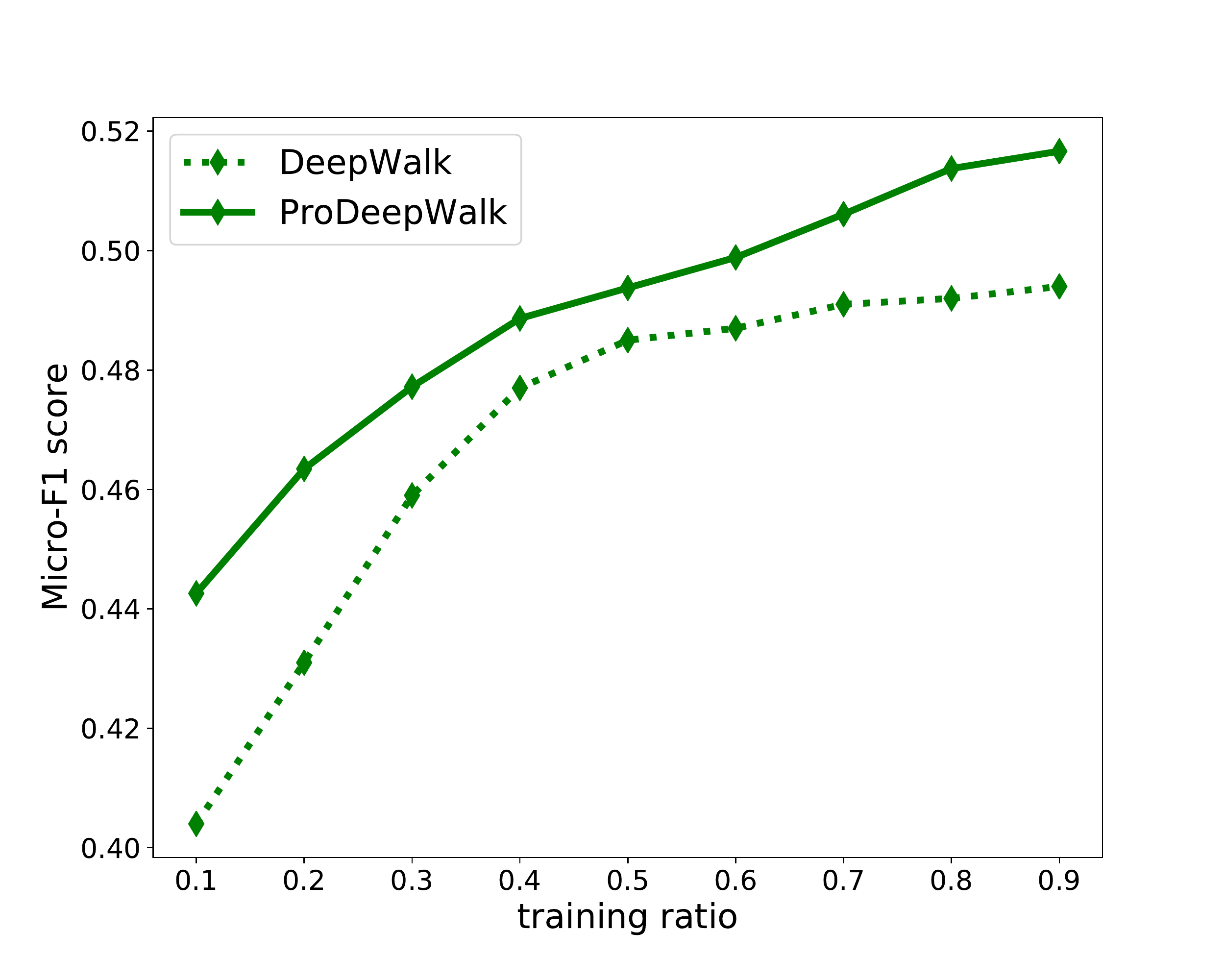}
\label{fig:ProDeepWalk}
}
\end{subfigure}
\begin{subfigure}[ProLINE]{
\centering
\includegraphics[width = 0.4 \linewidth]{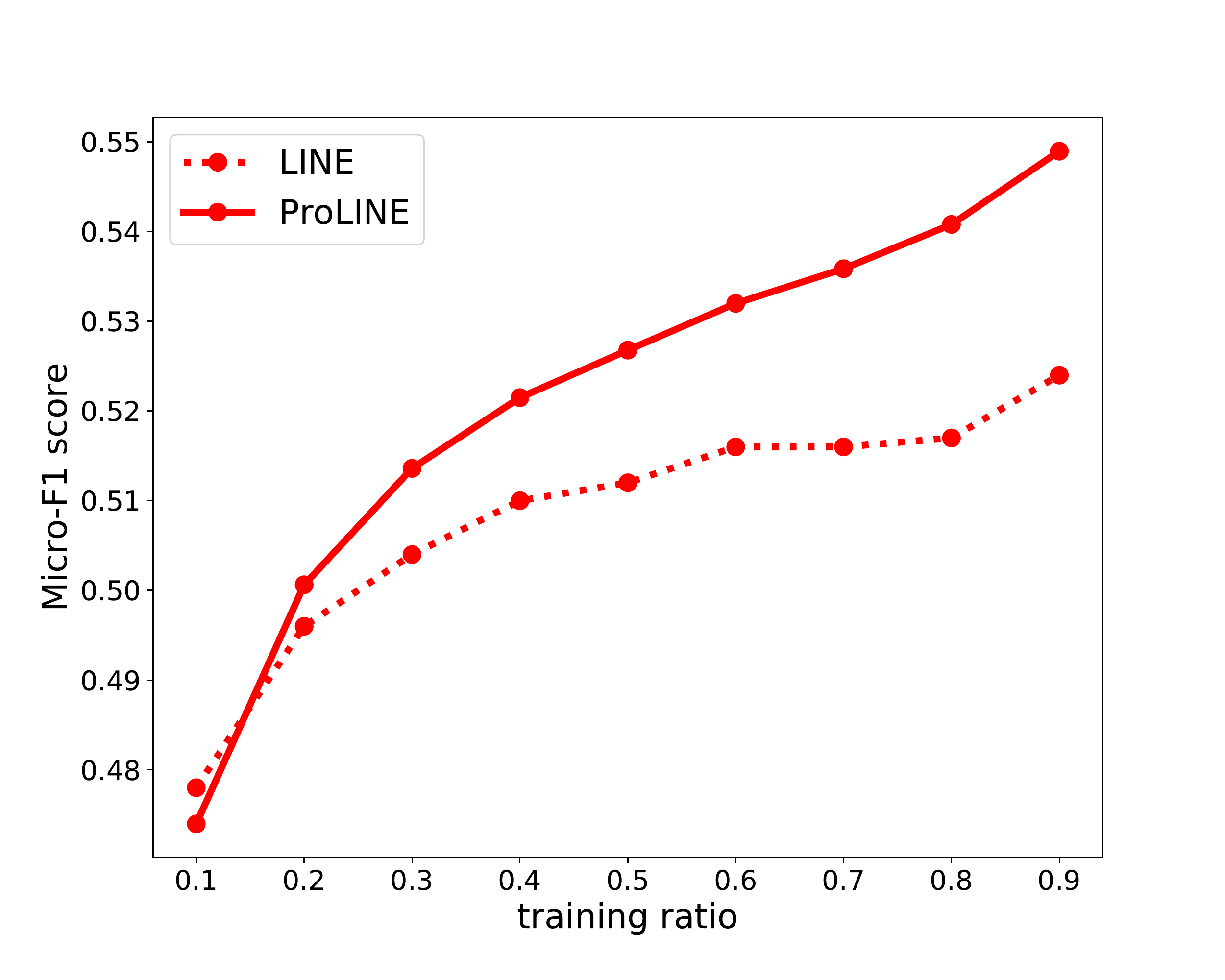}
\label{fig:ProLINE}
}
\end{subfigure}

\begin{subfigure}[Pronode2vec]{
\centering
\includegraphics[width = 0.4 \linewidth]{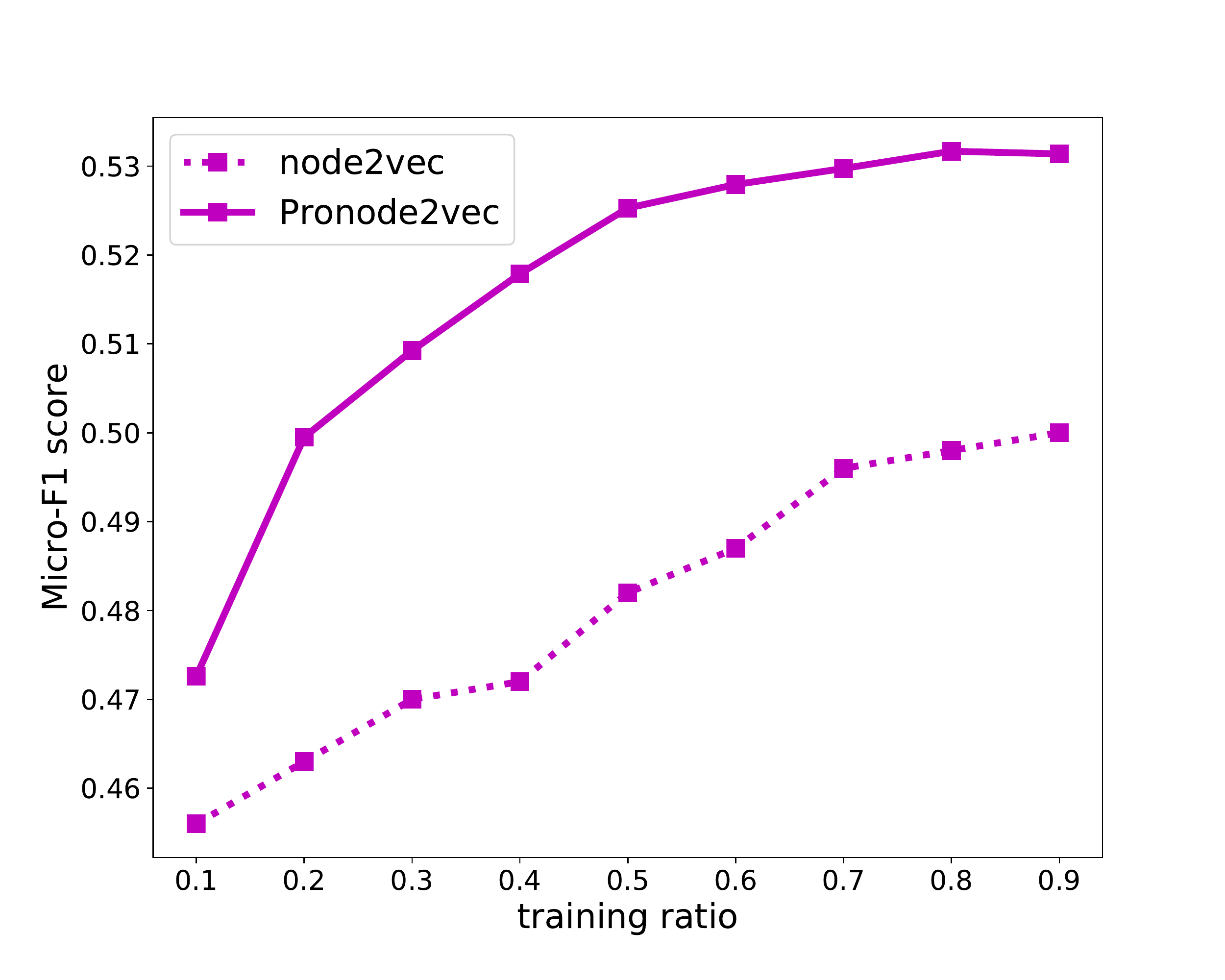}
\label{fig:Pronode2vec}
}
\end{subfigure}
\begin{subfigure}[ProHOPE]{
\centering
\includegraphics[width = 0.4 \linewidth]{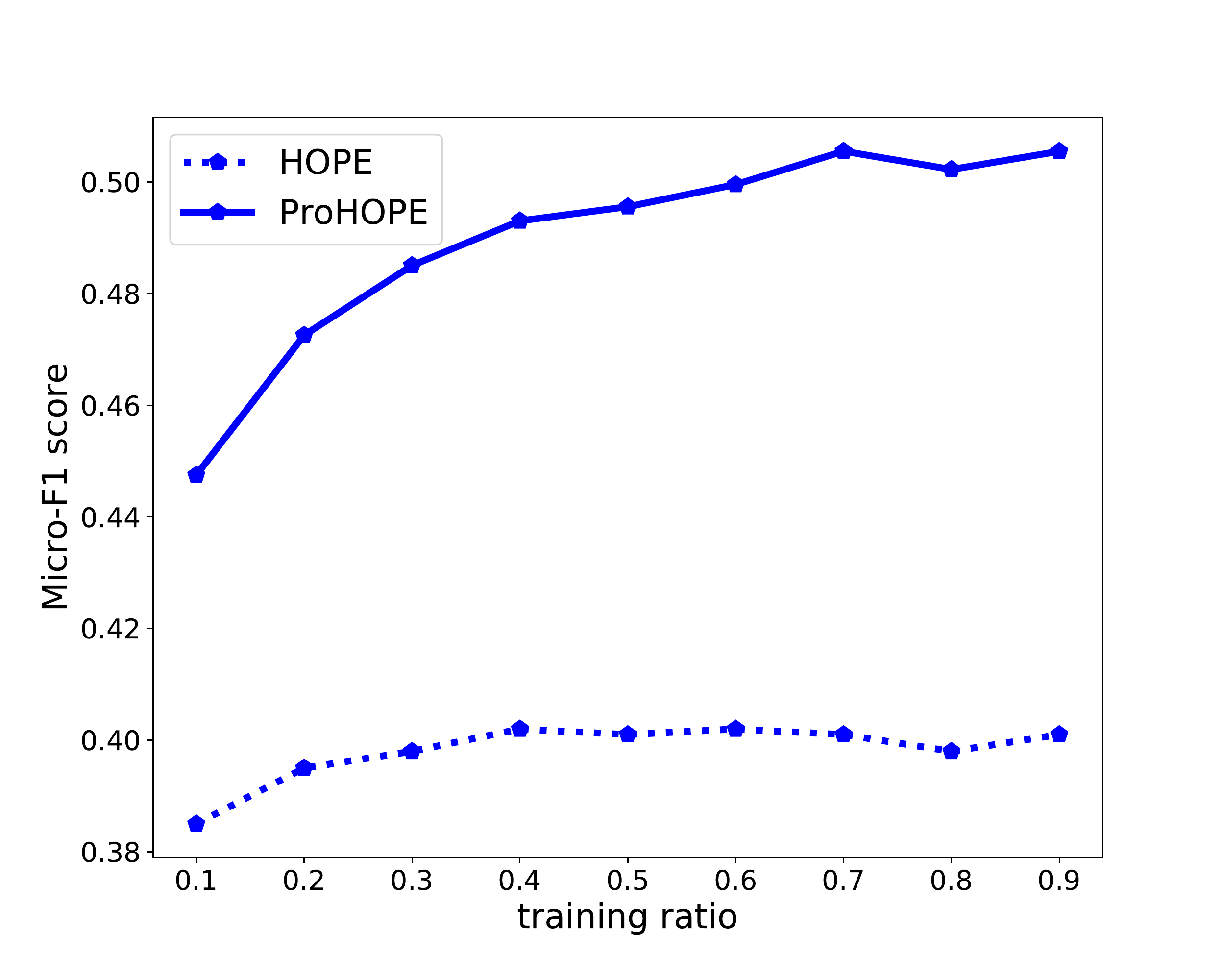}
\label{fig:ProHOPE}
}
\end{subfigure}
\caption{DeepWalk, LINE, node2vec and HOPE enhanced by Progle on Wikipedia}
\label{fig:ProBaseline}
\end{figure*} 

\subsubsection{Evaluation Methods}
For effectiveness validation, the prediction performance of  embedding is evaluated via the average Micro-F1 in the multi-label classification task, which has also been used in the works of Deepwalk, LINE, node2vec and GraRep. Following the same experimental procedure in Deepwalk, We randomly sample different percentages of the labeled nodes for training and the rest for testing. For PPI, Wikipedia, and Blogcatalog datasets, the training ratio is varied from $10\%$ to $90\%$. For DBLP, the training ratio is varied from $1\%$ to $9\%$.  For each comparison method, the resultant embeddings are used to train a one-vs-rest logistic regression classifier with L2 regularization. In the test phase, the one-vs-rest model yields a ranking of labels rather than an exact label assignment. To avoid the thresholding effect~\cite{tang2009large}, assume the number of labels for test data is given~\cite{tang2009large, perozzi2014deepwalk}. We repeat each experiment for $10$ times and report the average Micro-F1 and standard deviation. Analogous results hold for Macro-F1 or other accuracy metrics, and thus are not shown.

In the efficiency and scalability experiments, the efficiency is evaluated by the CPU time the method takes to get the embedding. The scalability of \smodel is analyzed by the time cost in the multiple scale and different density networks. 

The experiments were conducted on a Red Hat server with four Intel Xeon(R) CPU E5-4650 (2.70GHz) and 1T RAM.

\subsubsection{Parameter Settings}
To facilitate comparison of methods, we set the dimension of all the embedding $d=128$. For DeepWalk and node2vec, we follow Deepwalk's preferred parameters -- windows size $m=10$, walks per node $r = 80$, walk length $l = 40$. $p, q$ in node2vec are searched over \{0.25, 0.50, 1, 2, 4 \}. For LINE, learning rate $\alpha = 0.025$, negative sample number $k = 5$.
 For GraRep, the dimension of the concatenated embedding is $d=128$ for fairness. For HOPE, $\beta$ is calculated in authors' code, and searched over $(0,1)$ to improve performance. For \model, the term number of the Chebyshev expansion is $k=10$  and $\mu=0.1$, $\theta=0.5$ are the default.
    
\subsection{Multi-label classification on different datasets}

Table \ref{tab:Random Walk} summarizes the prediction performance of all methods on the four datasets. It shows that \smodel achieves  significant improvement over baselines on different datasets at different training ratios. As for Blogcatalog, \smodel only achieve relatively marginal improvement over Deepwalk and Node2vec, but with $>100\times$ efficiency (in figure~\ref{fig:Protime}). 

Combined with the efficiency experiment, we can conclude that  \smodel outperforms or in some case is comparable to these state-of-the-art comparison methods while is $10\times \sim 100\times$ faster and takes up less computation resources.



\begin{figure*}[htbp]
\centering
\mbox
{
\begin{subfigure}[degree $=10$]{
\centering
\includegraphics[width = 0.4 \linewidth]{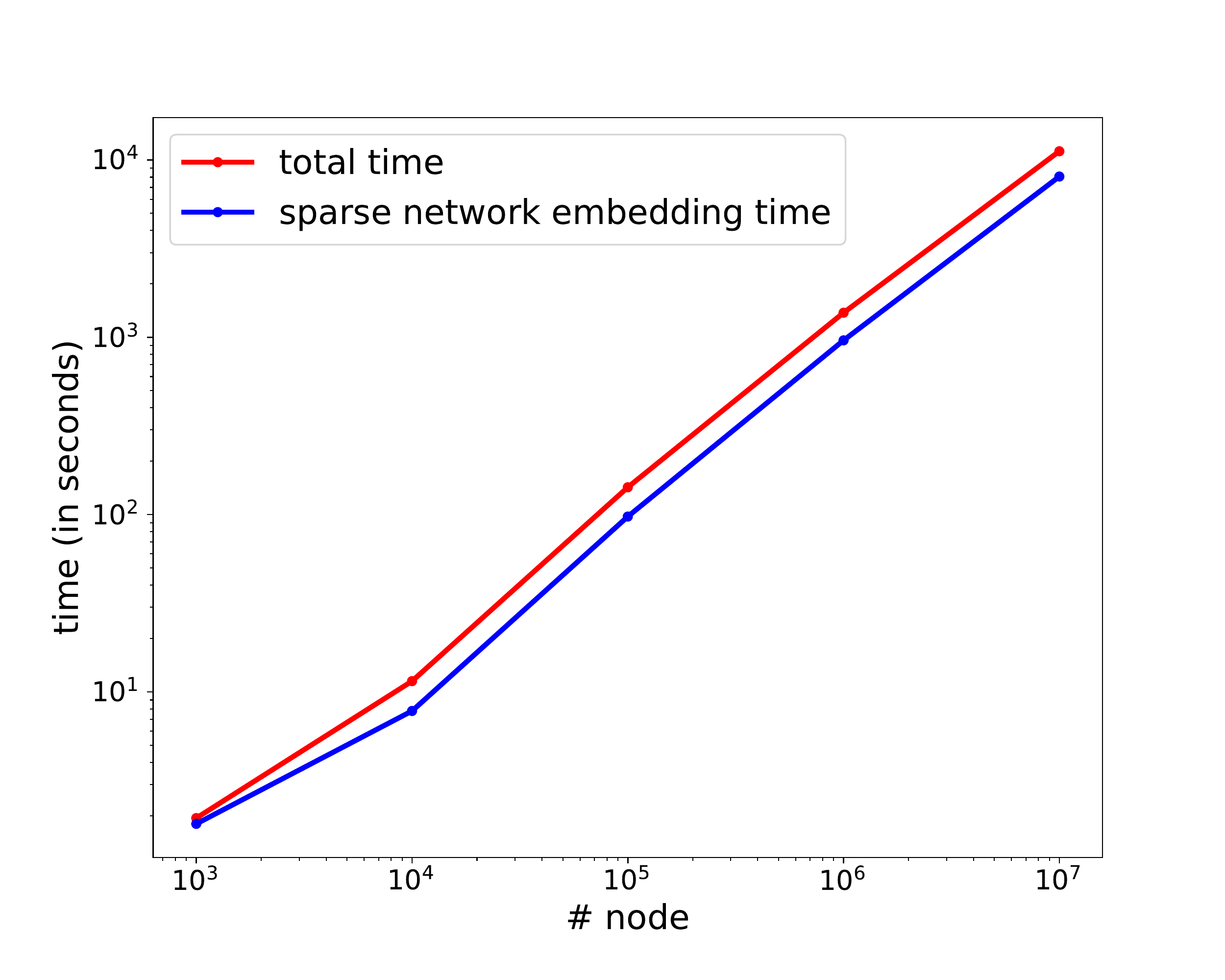}
\label{fig:ProDeepWalk}
}
\end{subfigure}
\hfill
\begin{subfigure}[\#node $=10000$]{
\centering
\includegraphics[width = 0.4 \linewidth]{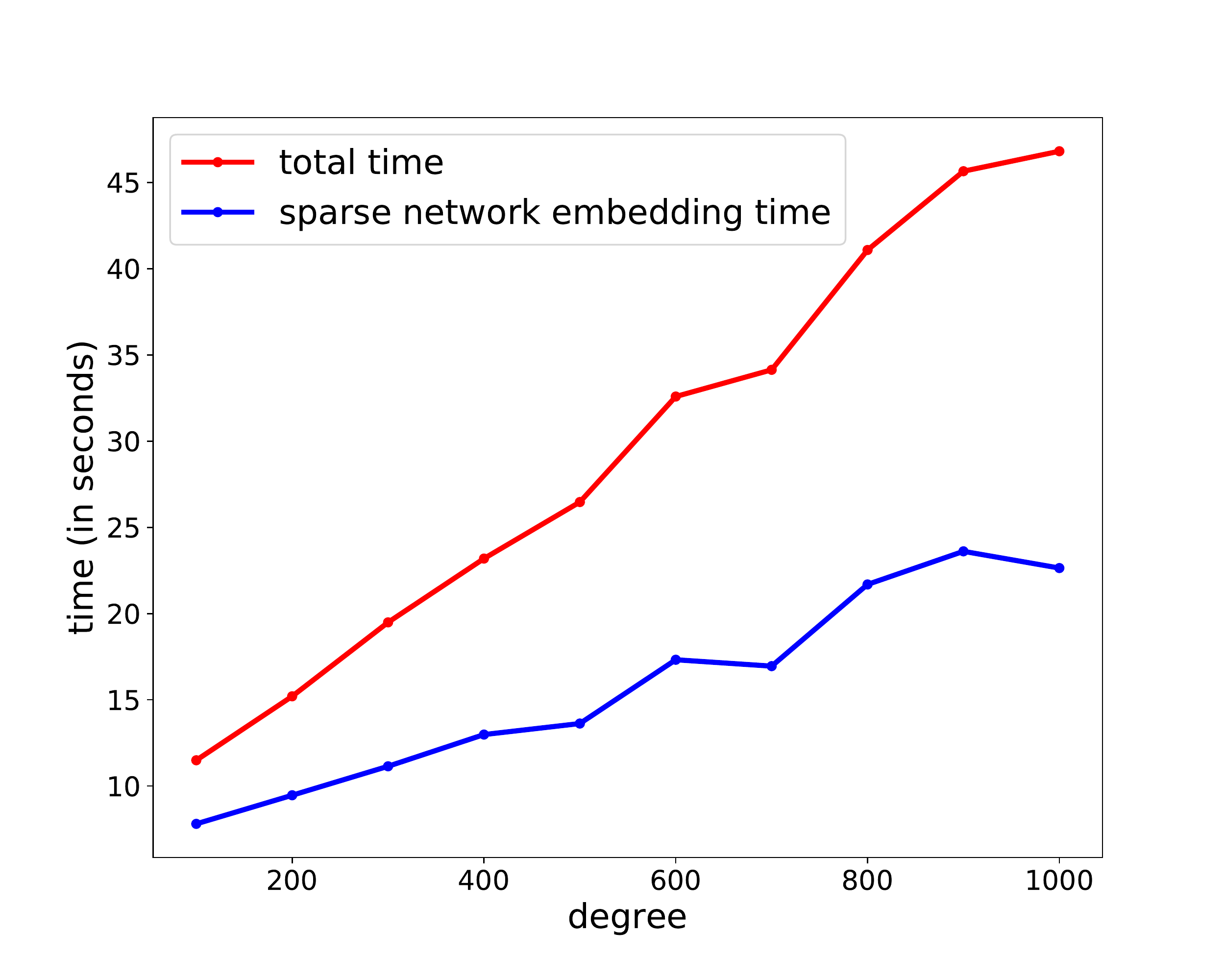}
\label{fig:ProLINE}
}
\end{subfigure}
}
\caption{Scalability analysis with respect to network volume and density}
\label{fig:scalability}
\end{figure*} 

\subsection{Enhancement for other embedding methods}
As we mentioned in the model section, spectral propagation is a general model to further incorporate network information and thus can enhance the quality of embedding trained by other models. 

On Wikipedia dataset, the accuracy performances of the baselines can be improved further. We input these learned embeddings into Progle and modulate them in the spectral space when propagated in the network. The outputs are denoted as ``Pro+'' embeddings. The improvement is significant (average $+9.4\%$ relatively gain for all methods at different training ratios; notably, $+26.2\%$ for HOPE at 80\% training ratio). The accuracy performances of ``Pro'' versions of these baselines are shown in Figure~\ref{fig:ProBaseline}.

These improvements are completed in a second, as the input embeddings are low-rank matrices ($n \times d$).

\begin{table}[htbp]
\centering
	\caption{Time cost statistics (/s).}
	\label{tab:time}
	\begin{tabular}{c c c c c}
\toprule[1.1pt]
		\textit{Dataset}  & \textit{PPI} & \textit{Wikipidia} & \textit{Blogcatalog} & \textit{DBLP} \\
		\hline
		\textit{Deepwalk} & 272 & 494  & 1231& 3825 \\
		\textit{LINE} & 70  & 87 & 185 & 1204\\
		\textit{node2vec} & 828 & 939 & 3533 & 4749 \\
		\textit{Progle} & \textbf{3} & \textbf{6} & \textbf{21} & \textbf{24} \\
		\bottomrule[1.1pt]
	\end{tabular}
\end{table}

\begin{table}[htbp]
\centering
	\caption{Time cost and scalability of Progle in  multiple scale networks (Single Thread).}
	\label{tab:time1}
	\begin{tabular}{c c c c}
\toprule[1.1pt]
		\textit{Dataset} & \#\textit{Nodes} & \#\textit{Edges} & \textit{time(/s)}   \\
		\hline
		\textit{Synthetic 1} & 10,000 & 500,000 & 11 \\
		\textit{Synthetic 2} & 10,000 & 1,000,000 & 15 \\
		\textit{Synthetic 3} & 10,000 & 1,500,000 & 19 \\
		\textit{Synthetic 4} & 10,000 & 2,000,000 & 23 \\
		\textit{Synthetic 5} & 10,000 & 2,500,000 & 26 \\
		\textit{Synthetic 6} & 10,000 & 3,000,000 & 32 \\
		\textit{Synthetic 7} & 10,000 & 3,500,000 & 34 \\
		\textit{Synthetic 8} & 10,000 & 4,000,000 & 41 \\
		\textit{Synthetic 9} & 10,000 & 4,500,000 & 45 \\
		\textit{Synthetic 10} & 10,000 & 5,000,000 & 47 \\
\bottomrule[1.1pt]
		\textit{Synthetic A} & 1,000 & 5,000 & 2 \\
		\textit{Synthetic B} & 10,000 & 50,000 & 6 \\
		\textit{Synthetic C} & 100,000 & 500,000 & 142 \\
		\textit{Synthetic D} & 1,000,000 & 5,000,000 & 1,375 \\
		\textit{Synthetic E} & 10,000,000 & 50,000,000 & 11,199 \\
\bottomrule[1.1pt]
		\textit{Flickr} & 80,513 & 5,899,882 & 150 \\
		\textit{Youtube} & 1,134,890 & 2,987,624 & 1,132  \\
		\textit{wiki-topcats} & 1,791,489 & 25,447,873 & 2,255  \\
		\bottomrule[1.1pt]
	\end{tabular}
\end{table}

\subsection{Efficiency and Scalability Experiment}
In this part, we first compare the efficiency of \smodel with other scalable embedding methods, that is, Deepwalk, LINE, and node2vec. Then we show the time cost of \smodel in larger networks and also analyze its scalability to validate the time and space complexity analysis. Notice that only \smodel is single threaded while DeepWalk and node2vec are trained using 20 workers, and iterative epoch is set 1, \smodel is more efficient than we show.

Table~\ref{tab:time} displays the time cost of Progle and word2vec-based methods. Utterly different from other spectral matrix factorization embedding methods, which take up all the CPU processes automatically and run much slow, \smodel is single threaded, but still $10 \times \sim 100 \times$ faster than word2vec-based embedding methods. In a word, Progle  takes up fewer computation and is faster without loss of effectiveness performance.

The time cost of Progle on real large-scale networks and multiple scale and density synthetic regular networks is shown in Table~\ref{tab:time1} and Figure \ref{fig:scalability}. For synthetic networks $1 \sim 10$, the number of nodes is fixed as 10,000 and the degree grows from $100$ to $100$; for synthetic networks A, B, C, D and E, the degree is fixed as $10$ and the number of nodes varies from $1,000$ to $10,000,000$. Its time cost increases nearly linearly with the number of nodes and edges, validating Progle's scalability. Note that the time cost of \smodel in the largest networks in Table~\ref{tab:time1} is still comparable to that of baselines in much smaller networks (Table~\ref{tab:time}).

Figure \ref{fig:scalability} shows both the total time cost and sparse network embedding time cost increase with respect to network volume and density, while the rest time spent in spectral propagation phase is the gap between these two curves. Therefore, the time cost in both sparse network embedding phase and spectral propagation phase is linear to network volume.

As for other spectral matrix factorization embedding methods, e.g. GraRep, they cause memory error when the network nodes $|V| > 50,000$.

\section{Related Work}
Roughly, related works mainly fall into three categories: spectral matrix factorization embeddings, word2vec-based embeddings, and convolution-based embeddings. 

Some original spectral network embedding works were related to spectral dimension reduction methods like Isomap~\cite{tenenbaum2000global}, Laplacian Eigenmaps~\cite{belkin2001laplacian} and spectral clustering~\cite{yan2009fast}. 
These methods typically exploiting the spectral properties of graph matrices, especially,  Laplacian and adjacency matrices. Their computation involves dense matrix decomposition, and thus are time consuming due to the time complexity of at least quadratic to the number of nodes, which limits their application. 

Another series of network embedding works have arisen since DeepWalk~\cite{perozzi2014deepwalk} built a bridge between network representation and neural language model. LINE~\cite{tang2015line} preserves the first-order proximity and second-order proximity  of networks. Node2vec~\cite{grover2016node2vec} designs a biased random walk procedure to make a trade-off between homophily similarity and structural equivalence. These methods often base on sampling, and optimize a non-convex objective, resulting in sampling error and optimization computation cost.
 A matrix decomposition comprehension for these word2vec-based methods was also be proposed, e.g., GraRep~\cite{cao2015grarep}, NetMF~\cite{qiu2017network}, and HOPE~\cite{ou2016asymmetric}. Like original spectral methods, these matrix decomposition versions of word2vec-based methods are also expensive both in time and space, and not scalable.

The third category of embedding method is convolution-based embedding method, e.g., Graph Convolution Network (GCN)~\cite{kipf2016semi}, Graph Attention Network (GAT)~\cite{Velickovic:17GAT}, Message Passing Neural Networks (MPNNs)~\cite{gilmer2017neural} and GraphSage~\cite{hamilton2017inductive}. In these works, convolution operation is defined in the spectral space and parametric filters are learned via back-propagation algorithm. Most of them are also not easily scaled up to handle large networks. Different from them, our model features a band-pass filter incorporating  both spatial locality smoothing and global structure, and the coefficients are calculated via integral in advance. Furthermore, the remarkable distinctions are that our method is without side-information features, unsupervised and task-independent, and benefits a broader range of downstream applications.

\section{Conclusions}
In this paper, network sparsity is utilized in two ways. First, edge dropout technique is used to design the sparse proximity matrix and sparse truncated SVD can be applied to get the embedding. Secondly, all the matrix product only involves sparse matrix, e.g. in the Chebyshev expansion phase. Third, sparse storage makes large-scale network embedding possible. Besides, spectral propagation in \smodel incorporate local and global network structure into embeddings, which can not only help \smodel outperform or at least be comparable to state-of-the-art baselines, but also enhance other methods at speed. In addition to the innovation on the model, the efficiency and effectiveness of the proposed method are also very significant. The model tackles the poor scalability and expensive computation
of spectral methods and also takes up fewer computation resources compared with existing methods.

\bibliographystyle{ACM-Reference-Format}
\bibliography{sample-bibliography}

\end{document}